\documentclass[aps,pra,twocolumn,amsfonts,amssymb,amsmath,showpacs,
floatfix,nofootinbib,groupedaddress,superscriptaddress,citesort]{revtex4}
\usepackage{mathrsfs}
\usepackage{amsfonts}
\usepackage{amstext}
\usepackage{amsmath}
\usepackage{amssymb}
\usepackage{bm}
\usepackage{CJK}
\usepackage{bbm}
\usepackage[dvips]{graphicx}
\def\qed{\leavevmode\unskip\penalty9999 \hbox{}\nobreak\hfill
     \quad\hbox{\leavevmode  \hbox to.77778em{%
              \hfil\vrule   \vbox to.675em%
               {\hrule width.6em\vfil\hrule}\vrule\hfil}}
     \par\vskip3pt}

\begin{document}

\title{Quantum correlation induced by the average distance between the reduced states}

\thanks{Dedicated to Prof. Jinchuan Hou, on the occasion of his
60th birthday}

\author{Yu Guo}
\affiliation{School of Mathematics
and Computer Science, Shanxi
Datong University, Datong 037009, China}%

\author{Xiulan Li}
\affiliation{School of Mathematics and Computer Science, Shanxi Datong University, Datong 037009, China}%

\author{Bo Li}
\affiliation{Department of Mathematics and Computer, Shangrao Normal University, Shangrao 334001, China}

\author{Heng Fan}
\affiliation{Institute of Physics, Chinese Academy of Sciences, Beijing 100190, China}

\begin{abstract}

A new quantum correlation in terms of the average distance between the
reduced state and the $i$-th output reduced states under
local von Neumann measurements is proposed. It is shown that only the product states do not
contain this quantum correlation and thus it is
different from both the quantum discord (QD) [Phys. Rev. Lett. \textbf{88}, 017901 (2001)] and the measurement-induced
nonlocality (MIN) [Phys. Rev. Lett. \textbf{106}, 120401(2011)]. For pure states,
it is twice of the quantity MIN, and is smaller than QD and entanglement of formation (EOF).
A general analytical formula is given and a lower bound for the two qubits case is obtained.
Furthermore, we compare it
with EOF and QD through
the Werner state and the isotropic state respectively.

\end{abstract}


\maketitle

\section{Introduction}

Quantum correlation not only is the key to our understanding
of quantum world but also is essential for the
powerful applications of quantum information and quantum
computation.
Entanglement lies at the heart of this field \cite{Horodecki,Guhne} but it does not
account for all possible quantum correlations contained in a bipartite system.
There are quantum correlations beyond entanglement, such as QD
\cite{Ollivier}, MIN \cite{Luo1},
measurement-induced disturbance (MID) \cite{Luo08,Long2013} and
measure of nonclassical correlation in coherence-vector representation \cite{Long2012}, etc.
QD has been proposed as the resource in quantum computation
\cite{Datta} and the Gaussian QD has beeen applied to quantum key distribution \cite{Suxiaolong2014}. It is indicated in
\cite{Luo1} that MIN may be useful in quantum cryptography, general
quantum dense coding \cite{Mattle}, remote state control
\cite{Bennett}, etc.

It is worth mentioning that, QD, MID and MIN are established by local measurement.
In addition, the use of averaged quantity in identifying quantum correlation has been used for
instance in quantifying entanglement for multipartite quantum systems \cite{Liudan2010,Cao2013}.
In this paper, we propose a new quantum correlation measured by the averaged quantity based on
local von Neumann measurements from a different point of view.
This paper is organized as follows. In Section 2, we give the definition of our quantum correlation.
Then Section 3 discusses the nullity of the our measure, it is showen that all quantum states
except the product one contain this kind of quantum correlation (Theorem 1).
Some numerical results are given in Section 4 (Propositions 2-4),
and in Section 5, three examples, i.e., the Bell-diagonal state,
the Werner state and the isotropic state are proposed and
we compare our measure with QD and EOF through the Werner state and the isotropic state.
At last we conclude in Section 6.

\section{Definition}

Let $H_a\otimes H_b$ with $\dim H_a=m$ and $\dim H_b=n\geq m$ be the state space of
the bipartite system A+B.
We now define
\begin{eqnarray}
Q(\rho):=\sup_{\Pi^a} \sum_k p_k\|\rho_b-\rho_b^{(k)}\|_2^2,
\end{eqnarray}
where $\|\cdot\|_2$ stands for the Hilbert-Schmidt norm (that is,
$\|A\|_2=[{\rm Tr}(A^\dag A)]^{\frac{1}{2}}$),
the supremum is
taken over all local von Neumann measurements $\Pi^a=\{\Pi_k^a\}$ (on the subsystem A),
$\rho_b^{(k)}=\frac{1}{p_k}
{\rm Tr}_a(\Pi_k^a\otimes I_b)\rho(\Pi_k^a\otimes I_b)$,
$p_k={\rm Tr}(\Pi_k^a\otimes I_b)\rho(\Pi_k^a\otimes I_b)$,
$\rho_b={\rm Tr}_a(\rho)$ is the reduced state of
$\rho$ and $I_b$ is the identity map on part B.

$Q(\rho)$ denotes the maximum `mean distance' between the local
state $\rho_b$ and the output local states $\rho_b^{(k)}$
with respect to the local measurement element $\Pi^a$.
This quantity can be measured only by local part after the
local von Neumann measurements of another part.
$Q(\rho)>0$ means that if part A performs a measurement, then
the local state of part B will change. Therefore
$Q$ define a quantum correlation since it cannot occurs in the classical world.
By the definition, it is obvious that $Q$ is
invariant under local unitary operation,
i.e.,
$Q(U_a\otimes U_b\rho U_a\otimes U_b)=Q(\rho)$
holds for any unitary operator
$U_{a}$ (resp. $U_b$) acting on $H_a$ (resp. $H_b$).

\section{The nullity of $Q$}

\smallskip

\noindent{\bf Theorem 1}\ {\it $Q(\rho)=0$ if and only if $\rho$ is a product state.}

\smallskip

\noindent{\it Proof} \ We only need to check the `only if' part.
If $Q(\rho)=0$, then for any local von Neumann measurement $\Pi^a=\{\Pi_k^a\}$,
we have $\rho_b^{(k)}=\rho_b$ for any $k$.
Let $\{|i\rangle\}$ and $\{|j'\rangle\}$ be the
orthonormal bases of $H_a$ and $H_b$, respectively.
Write $E_{ij}=|i\rangle\langle j|$.
Then any bipartite state $\rho$ can be written as
\begin{eqnarray*}
\rho=\sum_{i,j} E_{ij}\otimes B_{ij},\label{1}
\end{eqnarray*}
where $B_{ij}$s are operators acting on $H_b$.
It follows that
$B_{ii}\propto \rho_b$.
We show that $B_{ij}\propto \rho_b$ for any $i$ and $j$,
which guarantees that
$\rho$ is a product state.
With no loss of generality, we consider $B_{12}$.
Let $|\psi\rangle=\frac{1}{\sqrt{2}}(|1\rangle+|2\rangle)$
and $|\phi\rangle=\frac{1}{\sqrt{2}}(|1\rangle+i|2\rangle)$.
It turns out that the reduced states of the output states
under the local measurements
$|\psi\rangle\langle\psi|$ and $|\phi\rangle\langle\phi|$ are
\begin{eqnarray*}
\frac{1}{2}(B_{11}+B_{22}+B_{12}+B_{21})
\end{eqnarray*}
and
\begin{eqnarray*}
\frac{1}{2}(B_{11}+B_{22}+iB_{12}-iB_{21}),
\end{eqnarray*}
respectively.
This leads to $B_{12}+B_{21}\propto\rho_b$
and $B_{12}-B_{21}\propto\rho_b$, which reveals that
$B_{12}\propto\rho_b$
and $B_{21}\propto\rho_b$.
Therefore $\rho$ is a product state.
\hfill$\square$

\smallskip

Let $\rho=\sum_{i,j} A_{ij}\otimes E'_{ij}$,
where $E'_{ij}=|i'\rangle\langle j'|$. It is known that $\rho$ does not contain QD
if and only if $A_{ij}$s are mutually commuting normal
operators, and it does not contain MIN if and only if
$A_{ij}$s are mutually commuting normal operators and
each eigenspace of $\rho_a$ is contained in some eigenspace of
$A_{ij}$
for all $i$ and $j$ \cite{Guo3}.
Thus a state that does not contains QD or MIN may contains $Q$.

\section{Numerical Study}

\subsection{Pure state}

Let $|\psi\rangle$ be a pure state with
Schmidt decomposition $|\psi\rangle=\sum_k\lambda_k|k\rangle|k'\rangle$.
Let $\Pi^a=\{|e_i\rangle\langle e_i|\}$ be a von Neumann measurement on part A.
Write $\langle e_i|k\rangle=\alpha_{ik}$.
Then
\begin{eqnarray*}
\sigma_i&=&|e_i\rangle\langle e_i|\otimes I_b(|\psi\rangle\langle\psi|)|e_i\rangle\langle e_i|\otimes I_b\\
&=&\sum\limits_{k,l}\lambda_k\lambda_l|e_i\rangle\langle e_i|k\rangle\langle l|e_i\rangle\langle e_i|\otimes |k'\rangle \langle l'| \\
&=&|e_i\rangle \langle e_i|\otimes(\sum\limits_{k,l}\lambda_k\lambda_l\alpha_{ik}\bar{\alpha}_{il}|k'\rangle\langle l'|)\\
&=&p_i|e_i\rangle \langle e_i|\otimes|\psi_i\rangle\langle\psi_i|
=p_i|e_i\rangle \langle e_i|\otimes \rho_b^{(i)},
\end{eqnarray*}
where $|\psi_i\rangle=\frac{1}{p_i}\sum_k\lambda_k\alpha_{ik}|k'\rangle$,
$p_i=\|\sum_k\lambda_k\alpha_{ik}|k'\rangle\|^2=\sum_k\lambda_k^2|\alpha_{ik}|^2$.
It is straightforward that
\begin{eqnarray*}
\|\rho_b-\rho_b^{(i)}\|_2^2=2-\frac{2}{p_i}|\sum_k\alpha_{ik}\lambda_k^2|^2.
\end{eqnarray*}
Therefore
\begin{eqnarray*}
Q(\rho)&=&\sup\limits_{\Pi^a}\sum_ip_i(2-\frac{2}{p_i}|\sum_k\alpha_{ik}\lambda_k^2|^2)\\
&=&2-2\inf\limits_{\Pi^a}\sum\limits_i|\sum_k\alpha_{ik}\lambda_k^2|^2.
\end{eqnarray*}
Assume that the Schmidt rank of $|\psi\rangle$ is $r$.
Let $A=[\alpha_{ik}]$, $1\leq i\leq m$, $1\leq j\leq r$, and
let ${|\lambda\rangle}=(\lambda_1^2$, $\lambda_2^2$, $\dots$, $\lambda_r^2)^t$.
It follows that
\begin{eqnarray*}
\sum\limits_i|\sum_k\alpha_{ik}\lambda_k^2|^2&=&\|A{|\lambda\rangle}\|^2=\langle\lambda|A^\dag A|\lambda\rangle\\
&=&\langle\lambda|\lambda\rangle=\sum_i\lambda_i^4.
\end{eqnarray*}
We now get the following result.

\smallskip

\noindent{\bf Proposition 2}\ {\it Let $|\psi\rangle$ be a pure state with Schmidt coefficients $\{\lambda_i\}$ and Schmidt rank $r$.
Then
\begin{eqnarray}
Q(|\psi\rangle\langle\psi|)=2(1-\sum_i\lambda_i^4)\leq\frac{2(r-1)}{r},
\end{eqnarray}
and the equation holds if and only if $|\psi\rangle$ is maximally entangled.}

\smallskip

Let $N(\rho)$ denotes MIN of $\rho$ respectively.
Then  $N(|\psi\rangle\langle\psi|)=1-\sum_i\lambda_i^4$ \cite{Luo1}.
That is $Q(|\psi\rangle\langle\psi|)=2N(|\psi\rangle\langle\psi|)$.
The quantum discord of $|\psi\rangle$, $D(|\psi\rangle)$, which coincides with
the entanglement of formation $E$, is calculated as
$D(|\psi\rangle)=E(|\psi\rangle)=-\sum_i\lambda_i^2\log_2\lambda_i^2$.
Note that, for any $\lambda_i$, $\log_2\lambda_i^2-2\lambda_i^2\leq-2$,
we have
\begin{eqnarray}
Q(|\psi\rangle\langle\psi|)\leq D(|\psi\rangle)=E(|\psi\rangle).
\end{eqnarray}

\subsection{General case}

Let $\rho=\sum_i\delta_i E_i\otimes F_i$ be the operator Schmidt
decomposition of $\rho$ when $\rho$ is viewed as
a vector in the Hilbert space $\mathcal{B}(H_a)\otimes \mathcal{B}(H_b)$
with the Hilbert-Schmidt inner product
$\langle X|Y\rangle={\rm Tr}(X^\dag Y)$.
Let ${\rm Tr}(E_i)=\beta_i$ and ${\rm Tr}(F_i)=\gamma_i$.
Let $\{|k\rangle\}$ be an orthonormal basis of $H_a$.  Then
$\rho_b=\sum_i\delta_i\beta_i F_i$,
$p_k\rho_b^{(k)}=\sum_i\delta_i\alpha_{ki}F_i$, where
\begin{eqnarray}
\alpha_{ki}=\langle k|E_i|k\rangle.\label{2}
\end{eqnarray}
Therefore,
\begin{eqnarray*}
&&\sum_kp_k\|\rho_b-\rho_b^{(k)}\|_2^2
=\sum_kp_k\|\sum_i\delta_i(\beta_i-\frac{\alpha_{ki}}{p_k})F_i\|_2^2\\
&=&\sum_kp_k(\sum_i\delta_i^2(\beta_i-\frac{\alpha_{ki}}{p_k})^2)
=\sum_i\delta_i^2(\sum_k\frac{\alpha_{ki}^2}{p_k}-\beta_i^2).
\end{eqnarray*}
With the notations defined as above, we get the following theorem.

\smallskip

\noindent{\bf Proposition 3}\ {\it Let $\rho$ be a bipartite state with operator Schmidt coefficients
$\{\delta_i\}$, $T=[\alpha_{ki}]$ with $\alpha_{ki}$ defined as in Eq. (\ref{2}).
Then
\begin{eqnarray}
Q(\rho)=\sup_{T}\sum_i\delta_i^2(\sum_k\frac{\alpha_{ki}^2}{p_k}-\beta_i^2),
\end{eqnarray}
where the supremum is taken over all possible matrices $T=[\alpha_{ki}]$.}

\smallskip


\subsection{A lower bound for the two qubits case}

We now give a lower bound of the quantity $Q$ for the two qubits case.
Any two qubits state can be represented as
\begin{eqnarray}
\tau=\frac{1}{4}(I_a\otimes I_b+\vec{u}\vec{\sigma}\otimes I_b+I_a\otimes \vec{v}\vec{\sigma}+\sum\limits_{k,l=1}^3w_{kl}\sigma_k\otimes\sigma_l),
\end{eqnarray}
where
$\sigma_1=\left(\begin{array}{cc}0&1\\1&0\end{array}\right)$,
$\sigma_2=\left(\begin{array}{cc}0&-i\\i&0\end{array}\right)$ and
$\sigma_3=\left(\begin{array}{cc}1&0\\0&-1\end{array}\right)$
are the Pauli matrices, $\vec{u}=(u_1,u_2,u_3)$, $\vec{v}=(v_1,v_2,v_3)\in \mathbb{R}^3$, $w_{kl}$ are real numbers,
$\vec{\sigma}=(\sigma_1,\sigma_2,\sigma_3)$, and $\vec{u}\vec{\sigma}=\sum\limits_{i=1}^3u_i\sigma_i$, etc.
In fact, $\rho$ is locally unitary equivalent to \cite{Fano83,Luo2008}
\begin{eqnarray}
\rho=\frac{1}{4}(I_a\otimes I_b+\vec{a}\vec{\sigma}\otimes I_b
+I_a\otimes \vec{b}\vec{\sigma}+\sum\limits_{j=1}^3c_{j}\sigma_j\otimes\sigma_j),\label{qubits2}
\end{eqnarray}
where $c_j$s are real numbers.
Namely, there exists $2\times 2$ unitary matrices $U_a$ and $U_b$ such that
$\rho=U_a\otimes U_b\tau U_a^\dag\otimes O_b^\dag$.
That is, the state with the form in Eq.~(\ref{qubits2}) is enough when we discuss the quantum correlation $Q$
for the two qubits case since any quantum correlation is invariance under the local unitary operation.
Let $\Pi^a=\{\Pi_1^a,\Pi_2^a\}$,
$\langle k|\sigma_i| k\rangle=\alpha_{ik}$, $k=0$, 1.
Then $p_k\rho_b^{(k)}=\frac{1}{4}(I_b+\sum_ia_i\alpha_{ik}I_b+\vec{b}\vec{\sigma}+\sum_ic_i\alpha_{ik}\sigma_i)$,
$p_k=\frac{1}{2}(1+\sum_ia_i\alpha_{ik})$. Let $t_k=\frac{1}{1+\sum_ia_i\alpha_{ik}}$.
It turns out that
\begin{eqnarray*}
\|\rho_b-\rho_b^{(k)}\|_2^2&=&\frac{1}{2}\sum\limits_i|(1-t_k)b_i-t_kc_i\alpha_{ik}|^2.
\end{eqnarray*}
Therefore
\begin{eqnarray}
&&\sum\limits_kp_k\|\rho_b-\rho_b^{(k)}\|_2^2\nonumber\\
&=&\frac{1}{2}\sum\limits_k\frac{1}{t_k}\sum\limits_i|(1-t_k)b_i-t_kc_i\alpha_{ik}|^2.
\end{eqnarray}
Let $\Pi_1^a=|0\rangle\langle 0|$, $\Pi_2^a=|1\rangle\langle 1|$.
Then
\begin{eqnarray*}
\gamma_1&:=&\sum\limits_kp_k\|\rho_b-\rho_b^{(k)}\|_2^2\\
&=&\frac{1+a_1}{2}(|(1-\frac{1}{1+a_1})b_1-\frac{c_1}{1+a_1}|^2\\
&&+|(1-\frac{1}{1+a_1})b_2|^2+|(1-\frac{1}{1+a_1})b_3|^2)\\
&&+
\frac{1-a_1}{2}(|(1-\frac{1}{1-a_1})b_1+\frac{c_1}{1-a_1}|^2\\
&&+|(1-\frac{1}{1-a_1})b_2|^2+|(1-\frac{1}{1-a_1})b_3|^2).
\end{eqnarray*}
If $\Pi_1^a=|e_0\rangle\langle e_0|$, $\Pi_2^a=|e_1\rangle\langle e_1|$ with $|e_0\rangle=\frac{1}{\sqrt{2}}(|0\rangle+|1\rangle)$, $|e_1\rangle=\frac{1}{\sqrt{2}}(|0\rangle-|1\rangle)$,
then
\begin{eqnarray*}
\gamma_2&:=&\sum\limits_kp_k\|\rho_b-\rho_b^{(k)}\|_2^2\\
&=&\frac{1+a_2}{2}(|(1-\frac{1}{1+a_2})b_1|^2\\
&&+|(1-\frac{1}{1+a_2})b_2-\frac{c_2}{1+a_2}|^2+|(1-\frac{1}{1+a_2})b_3|^2)\\
&&+\frac{1-a_2}{2}(|(1-\frac{1}{1-a_2})b_1|^2
+|(1-\frac{1}{1-a_2})b_2\\
&&+\frac{c_2}{1-a_2}|^2+|(1-\frac{1}{1-a_2})b_3|^2).
\end{eqnarray*}
Similarly, if $\Pi_1^a=|f_0\rangle\langle f_0|$, $\Pi_2^a=|f_1\rangle\langle f_1|$ with $|f_0\rangle=\frac{1}{\sqrt{2}}(|0\rangle+i|1\rangle)$, $|f_1\rangle=\frac{1}{\sqrt{2}}(|0\rangle-i|1\rangle)$,
then
\begin{eqnarray*}
\gamma_3&:=&\sum\limits_kp_k\|\rho_b-\rho_b^{(k)}\|_2^2\\
&=&\frac{1+a_3}{2}(|(1-\frac{1}{1+a_3})b_1|^2
+|(1-\frac{1}{1+a_3})b_2|^2\\
&&+|(1-\frac{1}{1+a_3})b_3-\frac{c_3}{1+a_3}|^2)\\
&&+\frac{1-a_2}{2}(|(1-\frac{1}{1-a_3})b_1|^2
+|(1-\frac{1}{1-a_3})b_2|^2\\
&&+|(1-\frac{1}{1-a_3})b_3+\frac{c_3}{1-a_3}|^2).
\end{eqnarray*}
($\{|0\rangle,|1\rangle\}$, $\{|e_0\rangle,|e_1\rangle\}$ and $\{|f_0\rangle,|f_1\rangle\}$
are the eigenvectors of $\sigma_3$, $\sigma_1$ and $\sigma_2$, respectively.)
We now can conclude the following.

\smallskip

\noindent{\bf Proposition 4}\ {\it Let $\gamma=\max\{\gamma_i: i=1,2,3\}$, where $\gamma_i$s are defined as above, then
\begin{eqnarray}
Q(\rho)\geq\gamma.
\end{eqnarray}}

\smallskip

\section{Examples}

In general, $Q$ is hard to calculate for mixed states due to the supremum program.
However, it is easy for some special well-known states below.

\subsection{The two-qubit Bell-diagonal state}

We consider the two-qubit Bell-diagonal state
\begin{eqnarray}
\rho=\frac{1}{4}(I_a\otimes I_b+\sum\limits_{i=1}^{3}c_i\sigma_i\otimes\sigma_i),
\end{eqnarray}
where $c_i$s are real numbers, $\sigma_i$s are Pauli matrices, i.e.,
$\sigma_1=\left(\begin{array}{cc}0&1\\1&0\end{array}\right)$,
$\sigma_2=\left(\begin{array}{cc}0&-i\\i&0\end{array}\right)$ and
$\sigma_3=\left(\begin{array}{cc}1&0\\0&-1\end{array}\right)$.
\if false Let $\Pi_k^a=|\psi_k\rangle\langle\psi_k|$, $k=1$, 2, and
let $\langle\psi_k|\sigma_i|\psi_k\rangle=t_i^{(k)}$.
It turns out that
\begin{eqnarray*}
Q(\rho)&=&\frac{1}{2}(|\frac{1}{4}+c_3t_3^{(1)}|^2+|\frac{1}{4}-c_3t_3^{(1)}|^2\\
&&+|\frac{1}{4}+c_3t_3^{(2)}|^2+|\frac{1}{4}-c_3t_3^{(2)}|^2\\
&&+2|c_1t_1^{(1)}+c_2t_2^{(1)}i|^2+2|c_1t_1^{(2)}+c_2t_2^{(2)}i|^2).
\end{eqnarray*}
\fi
Let $c:=\max\{|c_1|,|c_2|,|c_3|\}$, use the same notations
as in \cite{Luo2008}, one can check that $\|\rho_b-\rho_b^{(1)}\|_2=\|\rho_b-\rho_b^{(2)}\|_2
=\|\frac{1}{2}(c_1z_1\sigma_1+c_2z_2\sigma_2+c_3z_3\sigma_3)\|_2
=\frac{1}{2}\sqrt{|c_1z_1|^2+|c_2z_2|^2+|c_3z_3|^2}$.
It follows that
\begin{eqnarray}
Q(\rho)=\frac{c^2}{4}.
\end{eqnarray}
That is, $Q(\rho)$ is decided only by the maximal value of $|c_i|$.




\begin{figure}
    \centering\includegraphics[width=8.5cm]{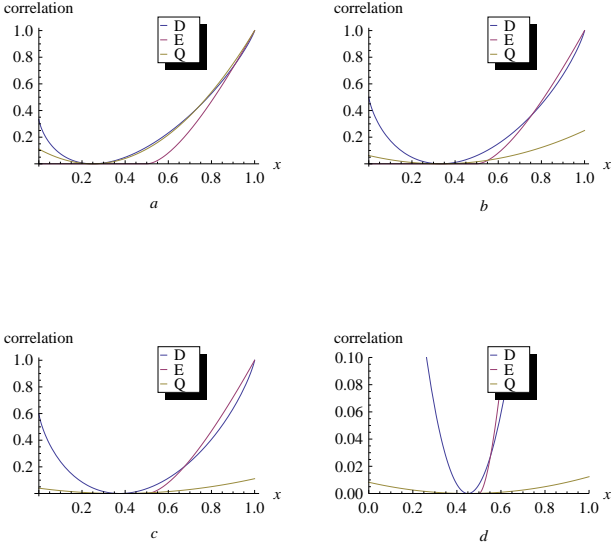}
    \caption{\label{fig: 1}(color online). $D$, $E$ and $Q$ in Werner state of various
dimensions with (a) $m=2$,(b) $m=3$,(c) $m=4$,(d) $m=10$.}
    \end{figure}

\subsection{The Werner state}

We consider the $m\otimes m$ Werner state
\begin{eqnarray}
\omega=\frac{2(1-x)}{m(m+1)}\Pi^++\frac{2x}{m(m-1)}\Pi^-,
\quad x\in[0,1], \label{p}
\end{eqnarray}
where $\Pi^+=\frac{1}{2}(I+F)$ and $\Pi^-=\frac{1}{2}(I-F)$ are projectors onto the
symmetric and antisymmetric subspace of $\mathbb{C}^m\otimes \mathbb{C}^m$ respectively,
$F=\sum_{i,j}|i\rangle\langle j|\otimes|j'\rangle\langle i'|$
is the swap operator.
For any von Neumann measurements $\Pi^a=\{|\psi_k\rangle\langle\psi_k|\}$ on part A,
let $\langle i|\psi_k\rangle=a_{ki}$.
Then
$\omega_b^{(k)}=\frac{m+2x-1}{m^2-1}I_b+\frac{m-2mx-1}{m^2-1}|\gamma_k\rangle\langle\gamma_k|$
with $|\gamma_k\rangle=\sum_ia_{ki}|i'\rangle$ and $p_k=\frac{1}{m}$ for any $1\leq k\leq m$.
Note that $\omega_b=\frac{1}{m}I_b$.
Therefore
\begin{eqnarray*}
\|\omega_b-\omega_b^{(k)}\|_2^2=\frac{(m-2mx-1)^2}{(m^2-1)^2},\label{3}
\end{eqnarray*}
which reveals that
\begin{eqnarray}
Q(\omega)=\frac{(m-2mx-1)^2}{(m^2-1)^2}.\label{4}
\end{eqnarray}
The equations above show that, for any local
von Neumann measurement $\Pi^a$ and for any element $\Pi^a_k$,
the distance between the local state and each output state
is a fixed constant and the probability of
each output state is $\frac{1}{m}$.


By the analytic formula of the quantum discord  and the
entanglement of formation (EOF) of the Werner state in~\cite{chitambar,werner},
we can compare $Q$ with QD ($D$) and EOF ($E$) (see in Figure 1).



\begin{figure}
    \centering\includegraphics[width=8.5cm]{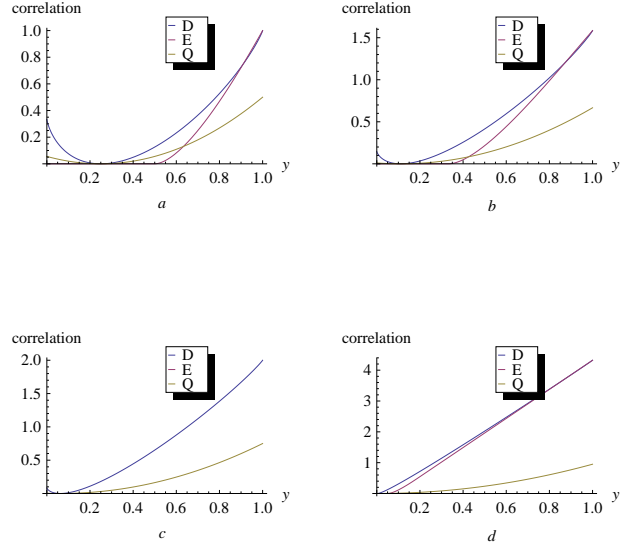}
    \caption{\label{fig: 2}(color online). $D$, $E$ and $Q$ in isotropic state of various
dimensions with (a) $m=2$,(b) $m=3$,(c) $m=4$,(d) $m=20$.}
    \end{figure}

\subsection{The isotropic state}


For the $m\otimes m$ isotropic state
\begin{eqnarray}
\varsigma=\frac{1-y}{m^2-1}I_a\otimes I_b
+\frac{m^2y-1}{m^2-1}|\psi^+\rangle\langle\psi^+|,\quad y\in[0,1],
\end{eqnarray}
where $|\psi^+\rangle$ is the maximally entangled state.
One can easily check that
\begin{eqnarray}
Q(\varsigma)=\frac{(m^2y-1)^2}{m(m+1)^2(m-1)}.
\end{eqnarray}
Combine with the analytic formula of quantum discord and
EOF for isotropic state in Ref.~\cite{chitambar}
and \cite{terhal} respectively, in Figure 2, we
compare $Q$ with QD and EOF for the isotropic states.

Note that, similar to the Werner state, for any local
von Neumann measurement $\Pi^a$ and for any element $\Pi^a_k$,
the distance between the local state and each local
output state is a fixed constant and the probability of
each output state is $\frac{1}{m}$ as well.

Figure 1 implies that, for the Werner state, the quantity $Q$ is lower than
$D$ for two qudits system with $d\geq3$, is incomparable with $D$ for two
qubits system and is incomparable with $E$ for system with any dimension in general.
Figure 2 implies that, for the isotropic state, the quantity $Q$ is lower than
$D$ and is incomparable with $E$.
That is, not only the nullity of $Q$ is different from $D$ and $E$,
the quantity of $Q$ is incomparable with that of $D$ and $E$ for some mixed sates.

\section{Conclusion}

We have established a new quantum correlation $Q$
according to local measurement with averaged local distance.
The nullity of this quantum correlation is shown to be
the set of all product states, namely, any non product
state contains quantum correlation.
A lower bound for two qubits case is proposed.
By comparing with entanglement and QD,
we find that the quantity $Q$ is quite different from both
QD and EOF.
It is remarkable that, by \cite{Rana}, although $Q(\omega)=N(\omega)$ and $Q(\varsigma)=\frac{m}{m-1}N(\varsigma)$,
but the nullity of $Q$ and $N$ does not coincide.
That is, all theses properties imply that $Q$ is a new kind of quantum correlation.
We hope that it maybe used in some quantum scenario based on local von Neumann measurements.

\begin{acknowledgements}
Y. Guo is supported by the Natural Science Foundation of
China (Grant No. 11301312, Grant No. 11171249) and the Natural Science Foundation of Shanxi
(Grant No. 2013021001-1,  Grant No. 2012011001-2).
B. Li is supported by the Natural Science Foundation of
China (Grant No. 11305015), the Natural Science Foundation of Jiangxi
Province (Grant No. 20132BAB212010).
H. Fan is
supported by the `973' program (Grant No. 2010CB922904).
\end{acknowledgements}



\end{document}